\documentclass[preprint,prb,aps,showpacs,groupedaddress,superscriptaddress]{revtex4-1}
\usepackage{graphicx}
\usepackage{amssymb}
\usepackage{textcomp}
\usepackage{amsmath}
\usepackage[normalem]{ulem} 
\usepackage{soul} 
\usepackage{gensymb}
\usepackage{textcomp}
\usepackage{amsmath}
\usepackage{siunitx}
\def\be{\begin{equation}}
\def\ee{\end{equation}}
\def\bea{\begin{eqnarray}}
\def\eea{\end{eqnarray}}
\def\bdm{\begin{displaymath}}
\def\edm{\end{displaymath}}
\def\ba{\begin{array}}
\def\ea{\end{array}}

\begin{document}

\title{Magnetic order and spin-orbit coupled Mott state in double perovskite (La$_{1-x}$Sr$_x$)$_2$CuIrO$_6$}

\author{W.~K.~Zhu}
\affiliation{Department of Physics, Indiana University, Bloomington IN 47405, USA}
\author{J.-C.~Tung}
\affiliation{Center for General Education, China Medical University, Taichung, 40402, Taiwan}
\author{W.~Tong}
\affiliation{High Magnetic Field Laboratory, Chinese Academy of Sciences, Hefei 230031, Anhui, P.~R. China}
\author{L.~Ling}
\affiliation{High Magnetic Field Laboratory, Chinese Academy of Sciences, Hefei 230031, Anhui, P.~R. China}
\author{M. Starr}
\author{J.~M. Wang}
\author{W.~C. Yang}
\affiliation{Department of Physics, Indiana University, Bloomington IN 47405, USA}
\author{Y. Losovyj}
\affiliation{Department of Chemistry, Indiana University, Bloomington IN 47405, USA}
\author{H.~D. Zhou}
\affiliation{Department of Physics and Astronomy, University of Tennessee, Knoxville TN 37996, USA}
\author{Y.~Q.~Wang}
\affiliation{High Magnetic Field Laboratory, Chinese Academy of Sciences, Hefei 230031, Anhui, P.~R. China}
\author{P.-H.~Lee}
\affiliation{The Affiliated Senior High School of National Taiwan Normal University, Taipei 10658, Taiwan}
\author{Y.-K.~Wang}
\affiliation{Center for General Education, National Taiwan Normal University, Taipei 106, Taiwan}
\author{Chi-Ken~Lu}
\email{Lu49@ntnu.edu.tw (theory)}
\affiliation{Physics Department, National Taiwan Normal University, Taipei 11677, Taiwan}
\author{S.~X.~Zhang}
\thanks{sxzhang@indiana.edu (experiment)}
\affiliation{Department of Physics, Indiana University, Bloomington IN 47405, USA}

\date{\today}

\begin{abstract}

Double-perovskite oxides that contain both 3d and 5d transition metal elements have attracted growing interest as they provide a model system to study the interplay of strong electron interaction and large spin-orbit coupling (SOC). Here, we report on experimental and theoretical studies of the magnetic and electronic properties of double-perovskites (La$_{1-x}$Sr$_x$)$_2$CuIrO$_6$ ($x$ = 0.0, 0.1, 0.2, and 0.3). The undoped La$_2$CuIrO$_6$ undergoes a magnetic phase transition from paramagnetism to antiferromagnetism at T$_N$ $\sim$ 74 K and exhibits a weak ferromagnetic behavior below $T_C$ $\sim$ 52 K. Two-dimensional magnetism that was observed in many other Cu-based double-perovskites is absent in our samples, which may be due to the existence of  weak Cu-Ir exchange interaction. First-principle density-functional theory (DFT) calculations show canted antiferromagnetic (AFM) order in both Cu$^{2+}$ and Ir$^{4+}$ sublattices, which gives rise to weak ferromagnetism. Electronic structure calculations suggest that La$_2$CuIrO$_6$ is an SOC-driven Mott insulator with an energy gap of $\sim$ 0.3 eV. Sr-doping decreases the magnetic ordering temperatures ($T_N$ and $T_C$) and suppresses the electrical resistivity. The high temperatures resistivity can be fitted using a variable-range-hopping model, consistent with the existence of disorders in these double-pervoskite compounds.

\end{abstract}

\maketitle

\section{Introduction}

Electron correlation and spin-orbit coupling (SOC) are the two key ingredients for realizing novel quantum phases,~\cite{1} such as spin-orbit coupled Mott insulators,~\cite{2,3,4} topological Mott insulators,~\cite{5,6} Weyl semimetals~\cite{7,8} and axion insulators.~\cite{9} 3d transition metals possess strong electron correlation because of their compact 3d orbitals but weak SOC due to their low atomic numbers; 5d metals, on the other hand, have much stronger SOC but weaker electron interactions.~\cite{2,3,10} The two groups of metals can be combined and form a rock-salt-like order in a double-perovskite oxide A$_2$BB'O$_6$, where A = rare earth element, B = 3d transition metal, and B' = 5d metal. These 3d-5d hybrid oxides have recently attracted growing interest because of their rich magnetic and electronic properties that arise from the interplay of the strong electron interaction and large SOC.~\cite{11,12,13,14,15,16,17,18,19,20,21,22,23,24,25,26,27,28,29,30,31,32,33,34,35,a,b,c,d,GChen1,GChen2} 

An important class of such double-perovskites are the iridates A$_2$BIrO$_6$ in which the electronic structure of iridium is strongly influenced by the SOC. Under the cubic symmetry, the 5d orbitals of iridium are split into $e_g$ and $t_{2g}$ states by the strong crystal field. The $t_{2g}$ state further forms a $J_{\rm {eff}}$=1/2 doublet and a $J_{\rm {eff}}$=3/2 quartet as a result of large SOC.~\cite{2,3} The five 5d electrons of Ir$^{4+}$ fills the $J_{\rm {eff}}$=3/2 state completely, leaving the $J_{\rm {eff}}$=1/2 state partially occupied. This $J_{\rm {eff}}$=1/2 picture, initially discovered in the layered Sr$_2$IrO$_4$,~\cite{2,3} has recently been demonstrated in the double perovskites La$_2$ZnIrO$_6$ and La$_2$MgIrO$_6$,~\cite{14} in which the B-site cations (i.e. Mg$^{2+}$ and Zn$^{2+}$) are non-magnetic. The Ir$^{4+}$ moments collectively form an antiferromagnetic (AFM) order in La$_2$MgIrO$_6$~\cite{14,36,37} but a strongly canted AFM in La$_2$ZnIrO$_6$,~\cite{14,16,36,37,38} in spite of their similar crystal structures. Recent neutron scattering experiment on both compounds along with theoretical calculations using a model Hamiltonian suggest the existence of dominant Kitaev interaction,~\cite{17} in addition to the superexchange couplings, and the Dzyaloshinskii-Moriya interactions.~\cite{16} 

Magnetic interactions and ordering become more complicated yet more interesting if the B-site cation is also magnetic (e.g., B = Co, Fe or Cu), in which case both the B-B and B-Ir interactions play important roles.~\cite{18,19,20,21,32,36,37,39} Most of the 3d magnetic cations have a variety of spin states (e.g., Co$^{2+}$ has low, medium and high spin states), which adds more complexity into the system. One of the exceptions is Cu$^{2+}$ which has nine 3d electrons with six of them filling the lower $t_{2g}$ states and the rest three filling the higher $e_g$ states. The degeneracy of the two $e_g$ orbitals (i.e. $d_{x^2-y^2}$ and $d_{z^2}$) can be lifted by the Jahn-Teller distortion of the CuO$_6$ octahedron. In the case where the $d_{x^2-y^2}$ orbital is half-filled and the $d_{z^2}$ is fully occupied, the magnetic interaction between the Cu$^{2+}$ moments is mainly in the corresponding x-y plane.~\cite{25} Such a two-dimensional magnetic behavior has been observed in some Cu-based double perovskites, including Sr$_2$CuWO$_6$, Sr$_2$CuMoO$_6$~\cite{25} and their doped compounds Sr$_2$Cu(W$_{1-x}$Mo$_x$)O$_6$.~\cite{22} 

Given the unique features of Ir$^{4+}$ and Cu$^{2+}$, it is of interest to study the double perovskite La$_2$CuIrO$_6$ that contains both elements. While the first La$_2$CuIrO$_6$ compound was synthesized back in 1965,~\cite{40} there have been very few studies over the past several decades~\cite{36,41,42} and its magnetic and electronic properties are far from being fully understood. Towards this end, we synthesized polycrystalline double-perovskite La$_2$CuIrO$_6$ along with its hole doped compounds (La$_{1-x}$Sr$_x$)$_2$CuIrO$_6$ ($x$ = 0.1, 0.2, and 0.3) and performed experimental and theoretical studies of their magnetic and electronic properties. The undoped La$_2$CuIrO$_6$ undergoes a magnetic phase transition from paramagnetism to antiferromagnetism at $T_N$ $\sim$ 74 K, and exhibit a weak ferromagnetic behavior below $T_C$ $\sim$ 52 K. Two-dimensional magnetism that was observed in many other Cu-based double-perovskites is absent in our samples. First-principle density-functional theory (DFT) calculations indicate that a canted-AFM order is formed in both Cu$^{2+}$ and Ir$^{4+}$ sublattices, giving rise to a weak FM component. Calculations of electronic properties provide an evidence that La$_2$CuIrO$_6$ is an SOC-driven Mott insulator. Sr-doping decreases the magnetic ordering temperatures ($T_N$ and $T_C$) as well as the electrical resistivity. 

\section{Experimental details and computational method}

Polycrystalline (La$_{1-x}$Sr$_x$)$_2$CuIrO$_6$ ($x$ = 0.0, 0.1, 0.2 and 0.3) samples were synthesized by conventional solid state reaction. High purity La$_2$O$_3$ (99.999\%) was first dried in air atmosphere at $\sim$ 900 $\celsius$  and was then mixed and ground with SrCO$_3$ (99.994\%), CuO (99.995\%) and IrO$_2$ (99.99\%) in stoichiometric ratios. The mixed powder was then heated in air at $\sim$ 900 $\celsius$ and $\sim$ 1050 $\celsius$ for a total of about 5 days with intermediate grindings. The samples were ground into powder for x-ray diffraction (XRD) characterization which was carried out using a PANalytical EMPYREAN diffractometer (Cu K$\alpha$ radiation). Rietveld refinements were performed on the XRD data using the GSAS software package.~\cite{43} As discussed in the Supplemental Data, the refinement of x-ray diffraction patterns suggests that the (La$_{1-x}$Sr$_x$)$_2$CuIrO$_6$ compounds have a monoclinic double-perovskite structure with the space group P21/n. Temperature dependent dc magnetic susceptibility and field dependent magnetization measurements were carried out in a Quantum Design Magnetic Property Measurement System with a superconducting quantum interference device (SQUID). The zero field cooled (ZFC) magnetization was taken upon warming in a magnetic field, and the field cooled (FC) magnetization was taken either upon warming or during cooling. Samples were pressed into pellets and then sintered for heat capacity and electrical resistivity measurements. The heat capacity characterizations were conducted in a Quantum Design Physical Property Measurement System. Electrical resistances were measured using either a Linear Research LR-700 ac Resistance Bridge or a home-built dc measurement system consisting of BNC-2110 connector block and DL1211 current preamplifier. The X-ray photoelectron spectroscopy (XPS) measurement was performed using a PHI VersaProbe II scanning XPS microprobe. As seen in the Supplemental data section, Cu is found to be in +2 state, and the majority of Ir ($\sim$ 95.42 \%) is in the +4 state along with a minor component ($\sim$ 4.58 \%) of +5 state. The existence of higher oxidation state may be attributed to slight non-stoichiometry, e.g. metal deficiency.   

In the fully relativistic first-principles calculations, we used the accurate frozen-core full-potential projector augmented-wave (PAW)~\cite{44, 45} method, as implemented in the Vienna Ab initio Simulation Package (VASP).~\cite{46, 47} The calculations were based on density functional theory, with the exchange and correlation effects being described by the generalized gradient approximation (GGA).~\cite{ 48, 49} The spin-orbit calculation was implemented in VASP by Kresse and Lebacq,~\cite{SOC1} where the fully relativistic augmented-plane-wave method was introduced by Kleinman~\cite{SOC2} and by MacDonald {\it et al.}.~\cite{SOC3} The relativistic Hamiltonian is given in a basis of total angular-momentum eigenstates which contain relativistic corrections up to $\alpha^2$,~\cite{SOC2} where $\alpha$ is the fine-structure constant. A very large plane-wave cutoff energy of 500 eV was used. The electronic and magnetic properties of La$_2$CuIrO$_6$ were calculated based on the experimental lattice parameters obtained from the Rietveld refinement of XRD pattern. Tests using optimized crystal structure suggest qualitatively similar electronic structures. The $\Gamma$-centered Monkhorst-Pack scheme with a k mesh of 6×4×5 in the full Brillouin zone (BZ), in conjunction with the tetrahedron method is used for the BZ integration. The total energy convergence criterion is 10$^{-6}$ eV/unit cell. To further take the d-electron correlation into account, we introduce the onsite Coulomb interaction U in the GGA+U approach with U = 5.0 (2.0) eV for Cu (Ir) are used and exchange parameter J = 0.9 (0.4) eV for Cu (Ir) are used. These correlation and exchange parameters are close to the typical values for 3d and 5d electrons in the literature.~\cite{25,c}

\section{Experimental Results}

\begin{figure}
\input{epsf}
\includegraphics[width=0.8\textwidth]{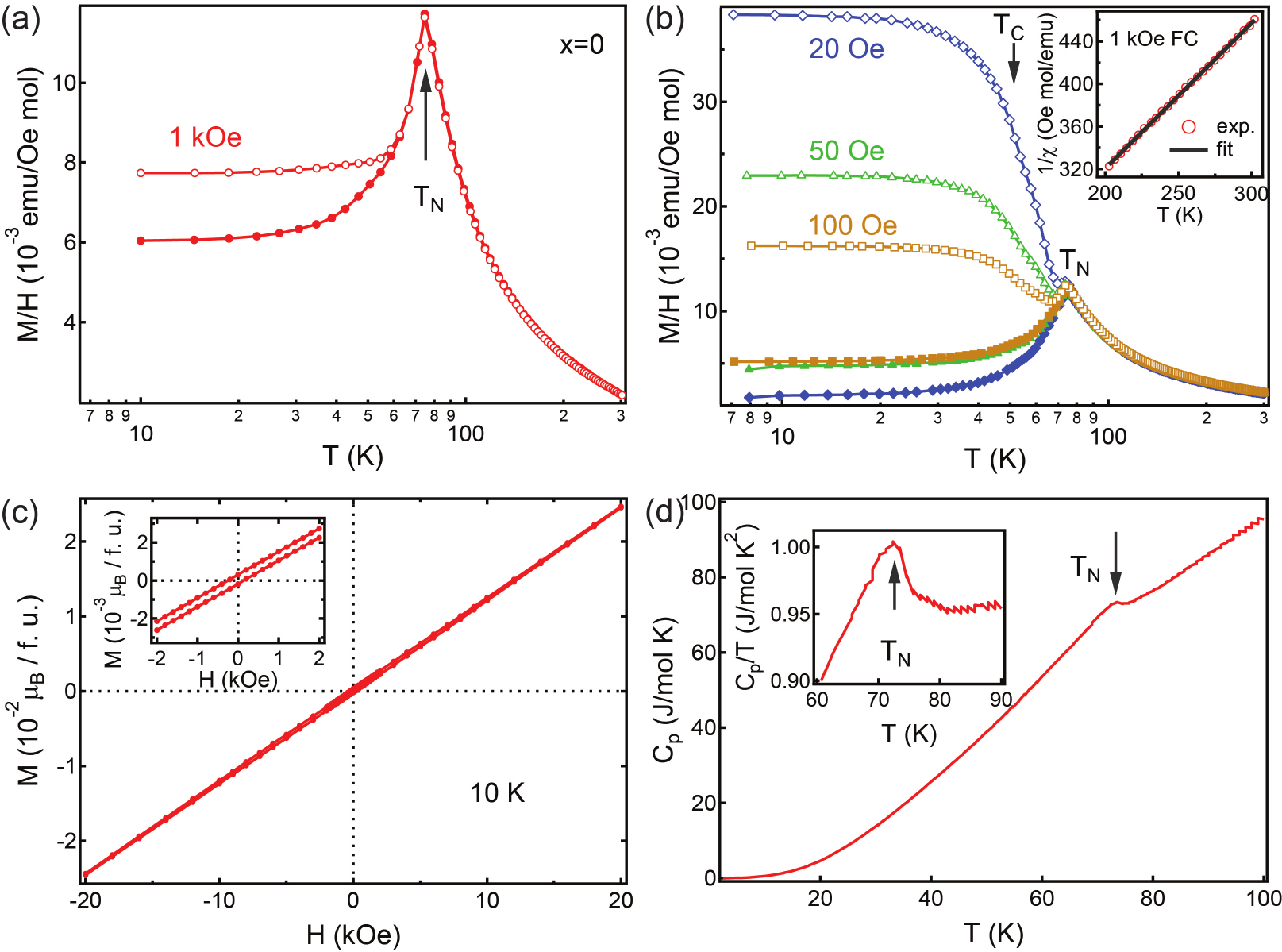}
\caption{(color online) Temperature dependence of ZFC (filled) and FC (open) magnetic susceptibilities taken on the undoped La$_2$CuIrO$_6$ at (a) 1 kOe and (b) lower fields (100, 50, and 20 Oe). Inset in (b): Curie-Weiss fitting to the data between 200 and 300 K. (c) Magnetic hysteresis loop of undoped La$_2$CuIrO$_6$ in a field range of -20 kOe to 20 kOe. Inset: M vs H at low fields. (d) Heat capacity of undoped La$_2$CuIrO$_6$. Inset: $C_p/T$ vs $T$ showing  transition temperature $T_N$.}\label{MT1}
\end{figure}

With the change of temperature, the undoped La$_2$CuIrO$_6$ undergoes two magnetic phase transitions. First, a paramagnetic to antiferromagnetic phase transition occurs at $T_N$ $\sim$ 74 K [Fig.~\ref{MT1}(a)], consistent with earlier studies.~\cite{36} This magnetic phase transition is confirmed by the temperature dependent heat capacity measurement, in which a peak is observed around $T_N$  [Fig.~\ref{MT1}(d) and its inset]. Second, when the temperature is further decreased below 60 K, a small hysteresis between the field-cooled (FC) and the zero-field-cooled (ZFC) magnetic susceptibilities appears, suggesting the existence of a weak FM component. The hysterisis becomes more pronounced when the magnetic field is decreased [Fig.~\ref{MT1}(b)]. This FM-like behavior is further confirmed by a small hysteresis loop in the field dependence of magnetization data taken at 10 K [Fig.~\ref{MT1} (c)].  By calculating $d\chi/dT$ as a function of temperature $T$, we determined the transition temperature $T_C$ of the weak FM to be about 52 K. The temperature dependence of heat capacity does not show a clear feature around $T_C$ [Fig.~\ref{MT1} (d)], indicating that the change in entropy is very weak at this ordering temperature. A Curie-Weiss fit to the high temperature magnetic susceptibility [inset of Fig.~\ref{MT1} (b)] gives rise to an effective moment $\mu_{\rm {eff,total}}$ = 2.416 $\mu_B$ and a Weiss constant $\theta$ = -34 K, indicating dominant AFM interactions. While two-dimensional order of Cu$^{2+}$ moments was observed in many Cu-based double perovskites A$_2$CuB'O$_6$ with B' cation of $d^0$ configuration,~\cite{22,25,51} it is not seen in our sample, manifested by the absence of a broad maxima in the temperature dependent susceptibility data [Fig.~\ref{MT1}(a) and (b)]. The absence of two dimensional magnetism may be due to some exchange couplings between Cu$^{2+}$ and Ir$^{4+}$. 

\begin{figure}
\input{epsf}
\includegraphics[width=0.8\textwidth]{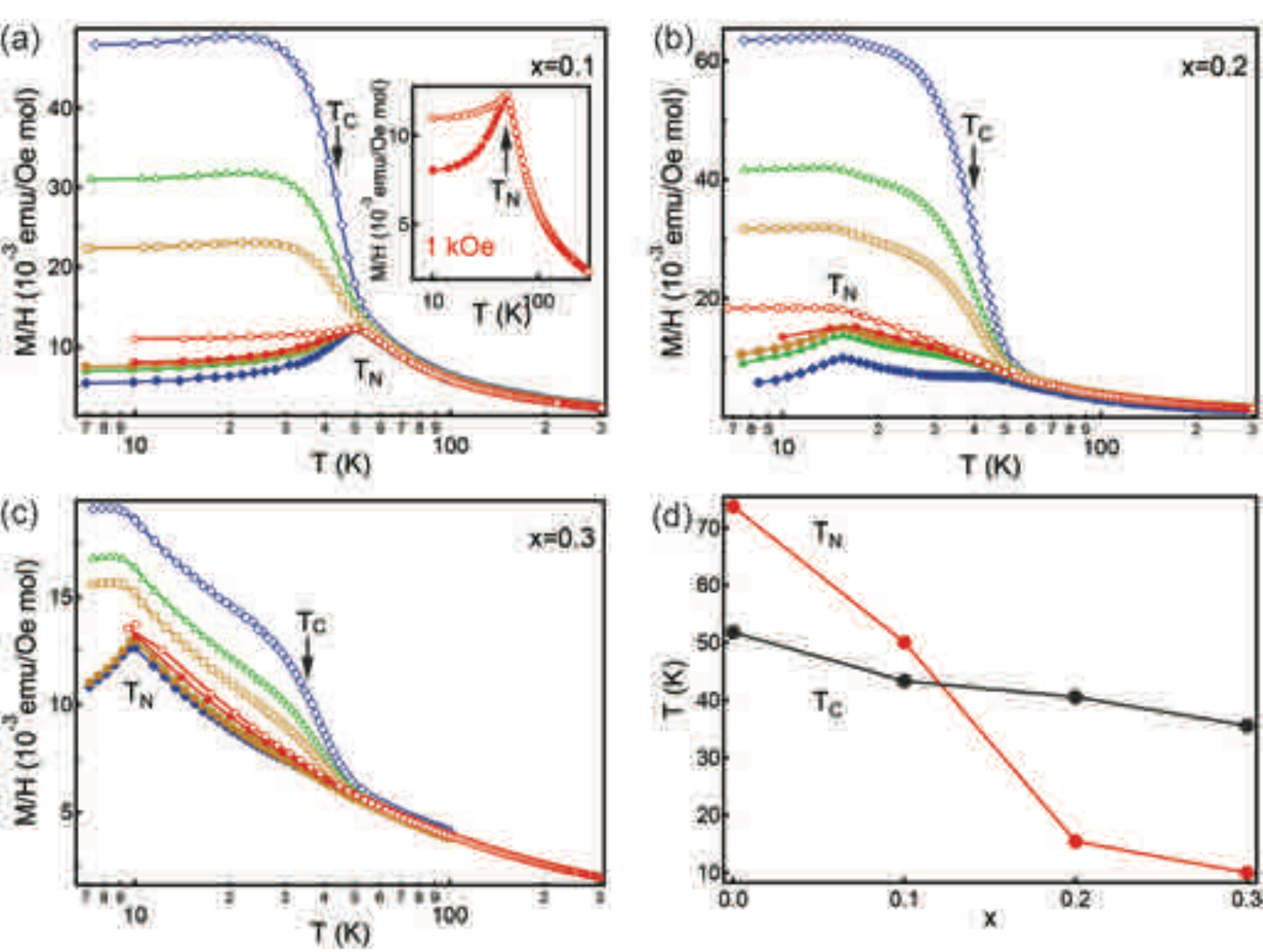}
\caption{(color online)Temperature dependence of ZFC (filled) and FC (open) magnetic susceptibilities taken on (a) $x$=0.1, (b) $x$=0.2, and (c) $x$=0.3 samples at different magnetic fields (1 kOe: red circles; 100 Oe: yellow squares; 50 Oe: green triangles and 20 Oe: blue diamonds). Inset in (a): $M$ vs $T$ curves taken at 1 kOe for $x$=0.1 sample. (d) Ordering temperatures $T_N$ and $T_C$ as a function of doping $x$.}\label{MT2}
\end{figure}

\begin{figure}
\input{epsf}
\includegraphics[width=0.4\textwidth]{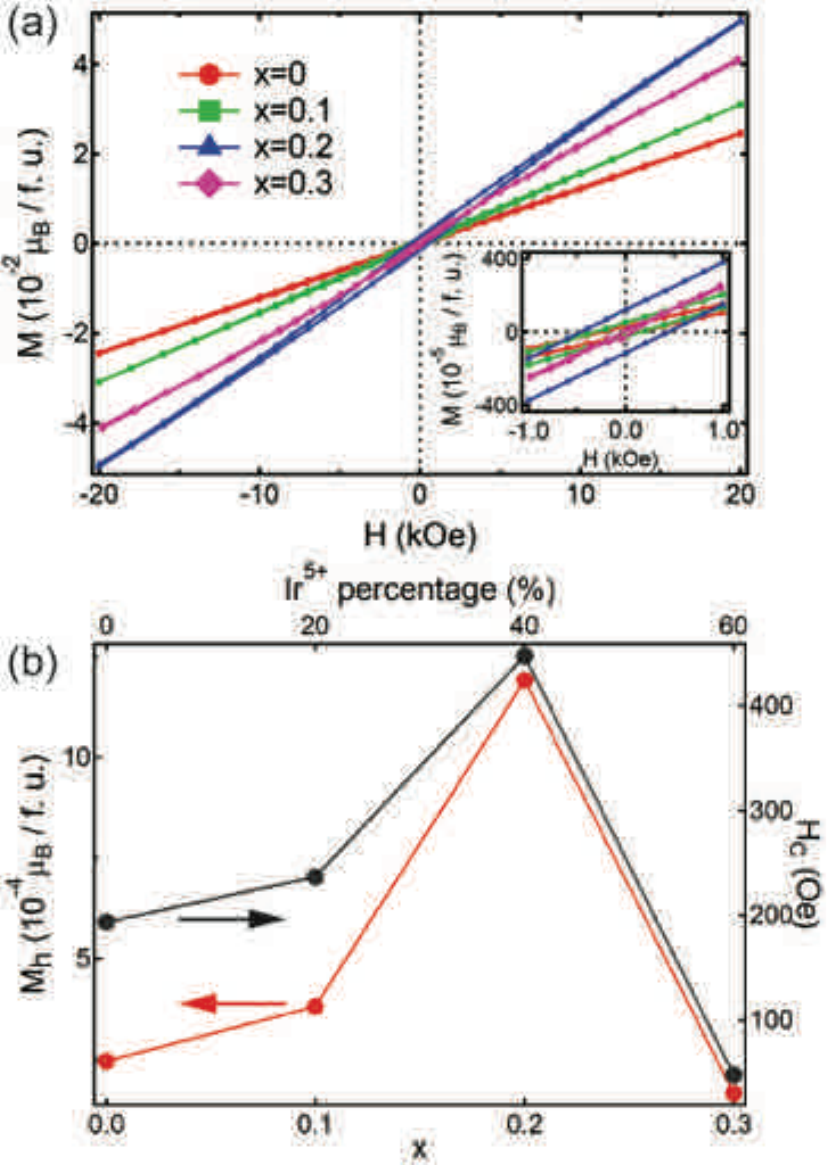}
\caption{(color online). (a) Magnetic hysteresis loops of (La$_{1-x}$Sr$_x$)$_2$CuIrO$_6$ samples in a field range of -20 kOe to 20 kOe. The inset shows the detailed hysteresis loops at low fields. (b) Remanent magnetization $M_h$ (left) and coercivity $H_C$ (right) as a function of doping level $x$ (bottom) and the percentage of Ir$^{5+}$ (top).}\label{MH}
\end{figure}

We further performed magnetic susceptibility measurement on the doped samples and observed a decrease in both $T_N$ and $T_C$. As shown in the inset of Fig.~\ref{MT2}(a), the susceptibility versus temperature for $x$ =0.1 sample shows a peak at $T_N\sim$ 50 K when the applied magnetic field is 1 kOe. Magnetic hystersis, as indicated by the divergence of ZFC and FC curves, appears at nearly the same temperature. Again the hystersis becomes clearer when the magnetic field is reduced [Fig.~\ref{MT2}(a)]. Using the $d\chi/dT$ versus $T$ data at 20 Oe, we determined the weak FM transition temperature $T_C$ to be $\sim$ 43 K. The magnetic susceptibility versus temperature data for $x$ =0.2 and 0.3 are plotted in Fig.~\ref{MT2}(b) and (c), respectively. The AFM ordering temperature $T_N$ is found to be lower than that of the weak FM ordering $T_C$. As seen more quantitatively in Fig.~\ref{MT2}(d), $T_N$ decreases more dramatically than $T_C$, and it reaches $\sim$ 10 K for $x$ =0.3. To study how the FM component changes with doping, we performed magnetization versus magnetic field (M-H) measurement at 10 K. As shown in Fig.~\ref{MH}(a), all of the M-H curves exhibit a small but visible hysteresis loop, accompanied by a linear field dependence of magnetization. Both the remanent magnetization ($M_h$) and the coercivity ($H_C$) increase when the doping is increased from $x$ =0 to $x$ =0.2 (i.e., the percentage of Ir$^{5+}$ increases from 0 to 40\%) [Fig.~\ref{MH}(b)]. Further increase of doping to 0.3 (i.e., Ir$^{5+}$/Ir = 60\%) results in a decrease of both the remanent magnetization and coercivity [Fig.~\ref{MH}(b)]. 

\begin{figure}
\input{epsf}
\includegraphics[width=0.8\textwidth]{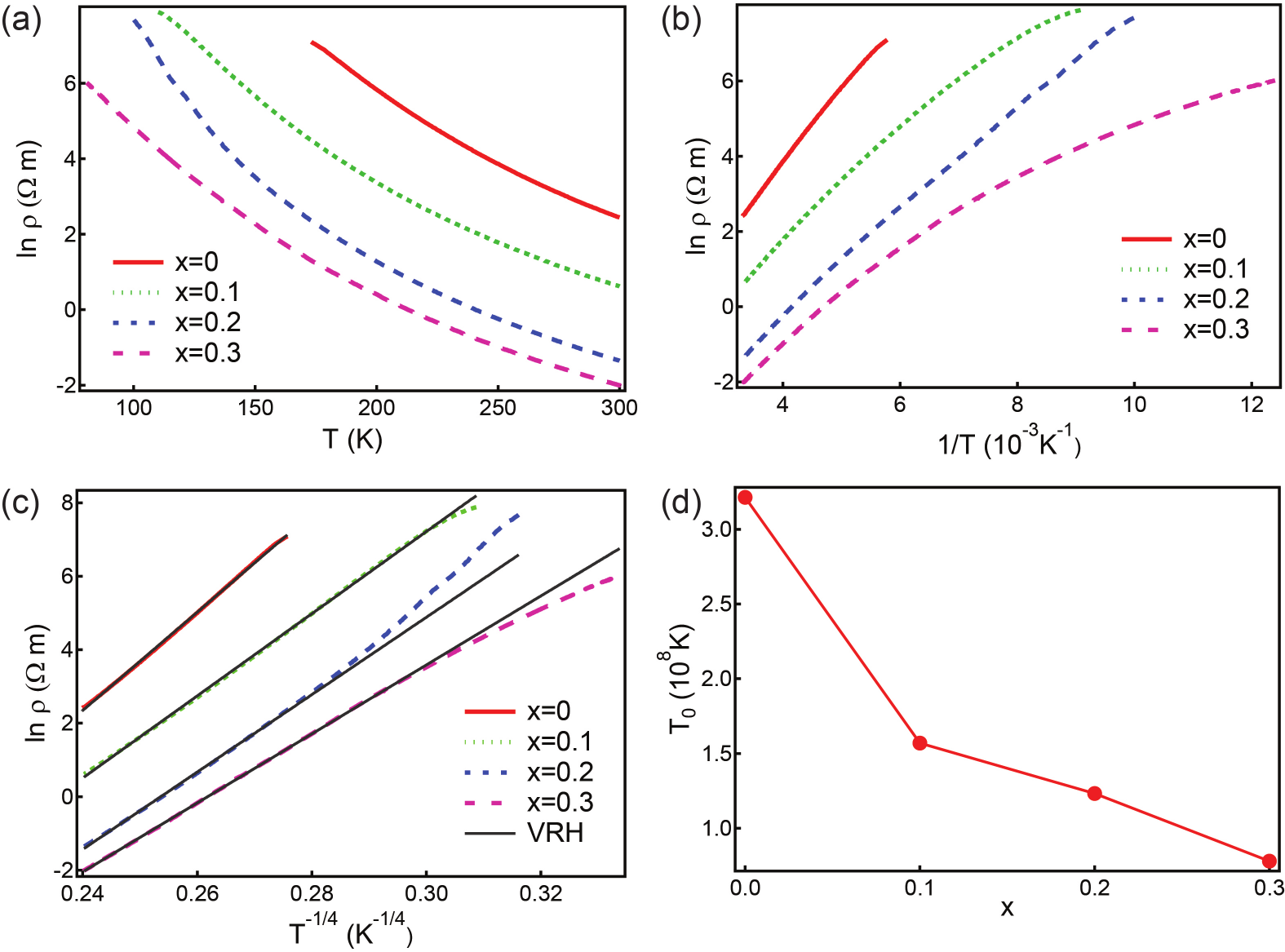}
\caption{(color online) (a) Electrical resistivity $\ln\rho$ vs $T$; (b) $\ln\rho$ vs $1/T$; and (c) $\ln\rho$ vs $T^{-1/4}$ for (La$_{1-x}$Sr$_x$)$_2$CuIrO$_6$. Black solid lines represent the VRH fittings. (d) $T_0$ as a function of doping $x$.}\label{RHO}
\end{figure}

An insulating behavior was observed in both the undoped and the doped (La$_{1-x}$Sr$_x$)$_2$CuIrO$_6$ ($x$ = 0.1, 0.2, and 0.3) samples, although Sr-doping decreases the electrical resistivity ($\rho$) significantly [Fig.~\ref{RHO}(a)]. While grain boundary scattering can certainly contribute to the high resistivity in our polycrystalline samples, it is unlikely to change significantly the temperature dependence.~\cite{grain} In fact, nearly all double perovskite oxides are found to be insulators~\cite{11} because of the large spatial separation between the B ions. Two different models were used to understand the electrical transport properties. The first one is the standard thermal activation model that is described by the equation $\rho(T)$=$\rho_0\exp(E_a/k_BT)$, where k$_B$ is the Boltzmann constant and $E_a$ is the activation energy. As shown in Fig.~\ref{RHO}(b), the undoped La$_2$CuIrO$_6$ shows a linear $\ln\rho$ versus $T^{-1}$ curve at high temperatures, while the doped samples exhibit a clear non-linearity. For the undoped sample,  an activation energy of  $E_a\sim$ 0.17 eV is determined from the slope of the $\ln\rho$ - $1/T$ curve. The second model is the disorder-induced variable-range-hopping (VRH) model,~\cite{56} in which $\rho(T)$=$\rho_0$ $\exp[(T_0/T)^{1/4}]$, where $T_0$ depends on the localization length $\ell$ and the density of available states $N(E)$ in a relation of $T_0$=18$/[N(E)\ell^3]$.~\cite{56,57,58} As shown in Fig.~\ref{RHO}(c), all of the $\ln\rho$ vs $T^{-1/4}$ curves are linear at high temperatures. The fact that the VRH transport persists near room temperature suggests considerable disorders in the samples, which is qualitatively consistent with the Rietveld refinements of XRD data. As shown in table S.1. the occupancies ($g_B$) of Cu and Ir on their correct sites are 0.87 to 0.92,  confirming the existence of certain degrees of disorder at the B-site. With the increase of doping $x$, the characteristic temperature $T_0$ decreases [Fig.~\ref{RHO}(d)], which may result from the increase in $N(E)$ as evidenced by the suppression of electrical resistivity [Fig.~\ref{RHO}(a)]. 

\section{DFT Calculations and discussions}

\begin{figure}
\input{epsf}
\includegraphics[width=0.8\textwidth]{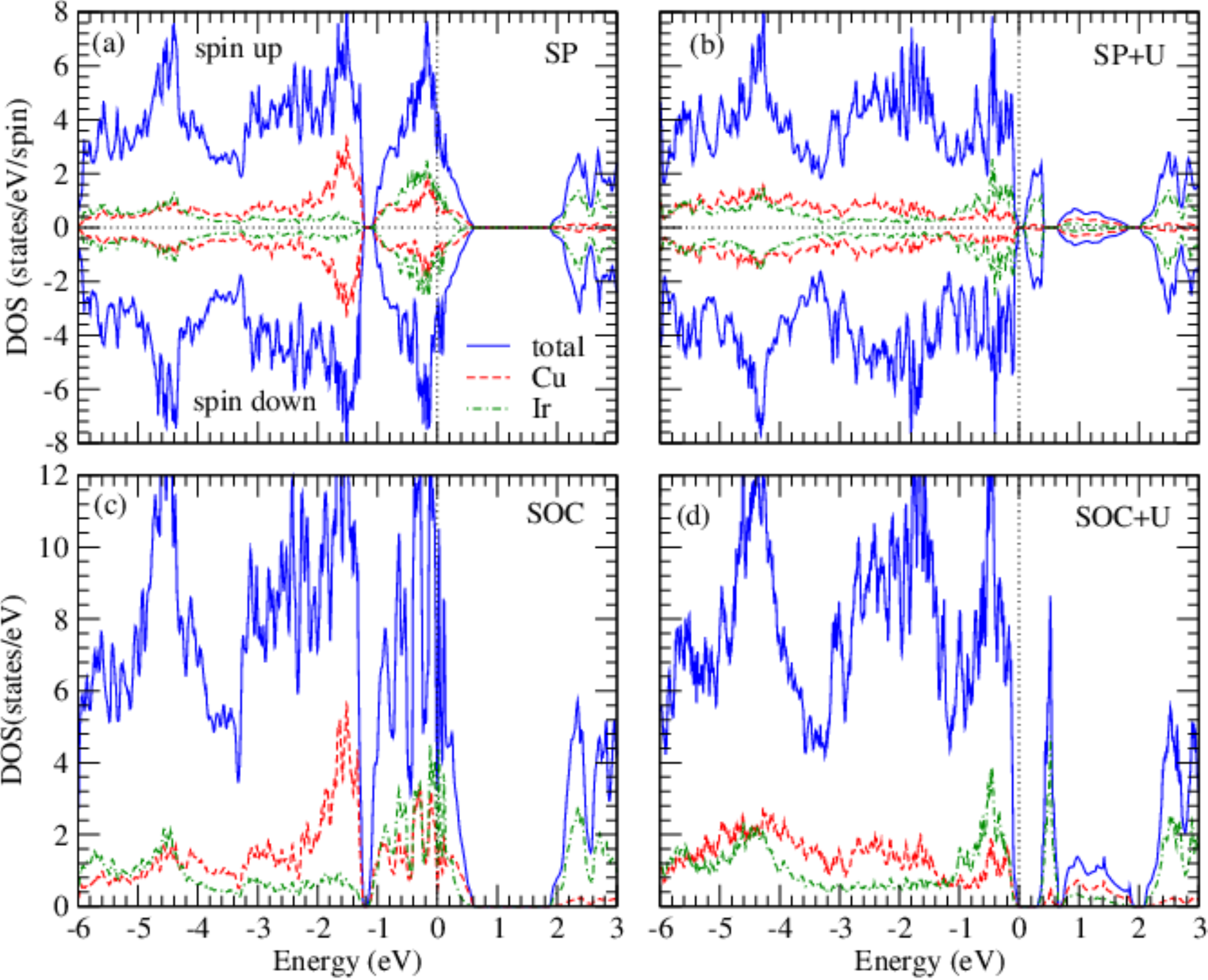}
\caption{(color online) The total and projected density of states of the undoped material La$_2$CuIrO$_6$ in four different situations: (a) the spin-polarized GGA calculation, (b) spin-polarized GGA+U, (c) GGA+SOC, (d) GGA+SOC+U. The Columb interactions U = 2 and 5 eV are adopted on Ir and Cu sites, respectively.}\label{DFT}
\end{figure}

DFT calculations of electronic properties suggest that the undoped La$_2$CuIrO$_6$ is an SOC driven Mott insulator. Fig.~\ref{DFT} displays the DOS for four different situations in calculations, namely the spin-polarized GGA calculation [Fig.~\ref{DFT}(a)], spin-polarized  GGA+U [Fig.~\ref{DFT}(b)], GGA+SOC [Fig.~\ref{DFT}(c)], and GGA+SOC+U [Fig.~\ref{DFT}(d)]. A sizeable gap of $\sim$ 0.3 eV is observed only in the case of GGA+SOC+U, which confirms that the measured insulating behavior is due to a combined effect of SOC and Coulomb interaction. The effect of U on the electronic structure can be seen clearly by comparing Fig.~\ref{DFT}(a) and (b). The upper Hubbard band of Ir is barely split off from the valence band, while the stronger $U_{\rm Cu}$ = 5 eV forces the upper Hubbard band of Cu to separate from the valence band. Hence, the GGA+U calculation with current parameters indicates an insulating state with zero gap.~\cite{footnote1} The absence of a finite band gap with reasonably large U values (5 eV for Cu and 2 eV for Ir) suggests that the electron correlation alone can not give rise to the insulating behavior observed in our experiment. A comparison between Fig.~\ref{DFT}(a) and {\bf (c)} manifests the effect of SOC: the otherwise broad band in the range of -1 to 0.5 eV [Fig.~\ref{DFT}(a)] seems to develope into a cluster of bands that have weak overlap with each other [Fig.~\ref{DFT}(c)]. The splitting of bands by SOC magnifies the effect of U, i.e. even a small $U_{\rm Ir}$ = 2 eV can open a Mott gap from a single narrow band [Fig.~\ref{DFT}(d)]. Our calculated results are hence qualitatively consistent with the $J_{\rm {eff}}$=1/2 picture demonstrated previously in Sr$_2$IrO$_4$.~\cite{2, 3} It is worth noting that the $J_{\rm {eff}}$=1/2 model is not necessary valid in all iridates. For example, a recent study suggests the breakdown of this model in another double-pervoskite iridate Sr$_2$YIrO$_6$.~\cite{Caod4} The DOS in Fig.~\ref{DFT} (d) also indicates weak energetic overlap between Cu and Ir. As a result, there could be a moderate exchange coupling between Cu$^{2+}$ and Ir$^{4+}$, which may be the reason for the absence of 2D magnetism discussed above.  

While DFT calculation of doped samples requires significantly enlarged unit cell and is out of the scope of the current work, the suppression of resistivity by Sr-doping can be understood within the $J_{\rm {eff}}$=1/2 picture. Replacing La$^{3+}$ by Sr$^{2+}$ introduces holes and changes Ir$^{4+}$ to Ir$^{5+}$. The empty $J_{\rm {eff}}$=1/2 level on the Ir$^{5+}$ site provides available states for the S=1/2 electron of Cu$^{2+}$ and the $J_{\rm {eff}}$=1/2 electron of Ir$^{4+}$ to hop, which gives rise to the suppression of resistivity. It is noted  that 5\% hole doping can induce an insulator-to-metal transition in pyrochlore Y$_2$Ir$_2$O$_7$,~\cite{54, 55} whereas the much more heavily doped double perovskite La$_{1.4}$Sr$_{0.6}$CuIrO$_6$ here still exhibits an insulating behavior. The insulating behavior of doped samples should be attributed to the long spatial distance between Ir$^{4+}$ and Ir$^{5+}$, the weak energetic overlap, and different orbital symmetries between $e_g$ of Cu$^{2+}$ and $t_{2g}$ of Ir$^{5+}$, which inhibits electron hopping as evidenced by the narrow band above the Fermi level.

\begin{figure}
\input{epsf}
\includegraphics[width=0.6\textwidth]{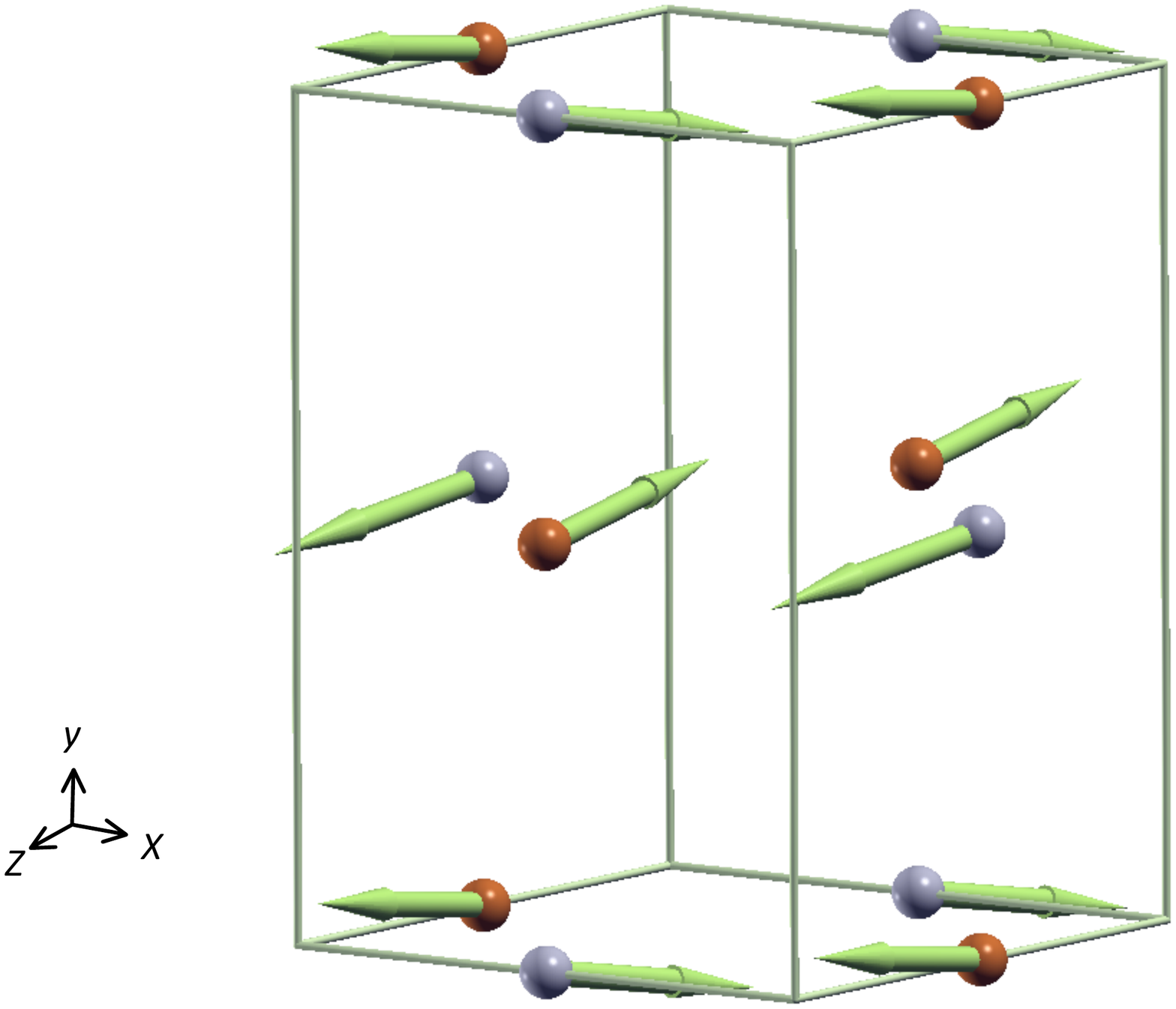}
\caption{(color online) Illustration of ordering of magnetic moments $\vec M$ on Cu$^{2+}$ (orange) and Ir$^{4+}$ (grey). A positive and negative net moment along y axis appear on Cu$^{2+}$  and Ir$^{4+}$ sublattices, respectively, leading to a total ferromagnetic moment 0.007 $\mu_B$ per f.u..}\label{AFM}
\end{figure}

DFT calculations of magnetic properties suggest that the ground state has a slightly canted antiferromagnetic order formed in both the Cu$^{2+}$ and the Ir$^{4+}$ sublattices. The magnetic structure is shown in Fig.~\ref{AFM}. The Cu$^{2+}$ and the Ir$^{4+}$ have a magnetic moment of $\sim$ 0.71 and 0.64 $\mu_B$, respectively, and the moments in the adjacent xz planes are nearly antiparallel to each other. As shown in Fig.~\ref{AFM}, a small ferromagnetic component along y axis can be observed on both Cu and Ir sites. The ferromagnetic components from Ir and Cu are opposite to each other but are not fully compensated, which leads to a small net moment of $\sim$ 0.007 $\mu_B$ per f.u.. The detailed spin and orbital moments of the two transition-metal cations in a doubled unit cell are summarized in Table~\ref{table1}. One should notice that the orbital moment $\vec L$ on the Ir site is not quenched due to the presence of strong SOC and it is in fact the major contribution to the total magnetic moment on Ir sites. The $\vec S$ and $\vec L$ in Ir$^{4+}$ are nearly parallel with each other (the angle between them is $\sim$ 2.87 \degree), which is another feature of the $J_{\rm {eff}}$=1/2 state.~\cite{2} In contrast, the spin moment $\vec S$ of Cu$^{2+}$ is much larger than its orbital moment, which is often the case in 3d cations. We note that the effective total moment $\mu_{\rm eff, total}$ = 2.416 $\mu_B$ obtained from the Curie-Weiss fitting is close to the ideal value $\sqrt{\mu_{\rm Cu}^2+\mu_{\rm Ir}^2}$ = 2.449 $\mu_B$ with $\mu_{\rm Cu}$ = $\mu_{\rm Ir}$ = 1.732 $\mu_B$ (The corresponding magnetic moments are 1 $\mu_B$ for both Cu$^{2+}$ and Ir$^{4+}$). Nevertheless, the DFT calculated values are lower than these ionic values, as observed in other cuprates~\cite{52} and iridates.~\cite{53} The reduction of moment may be attributed to the covalency between the metal d orbitals and the oxygen 2$p$ orbitals.~\cite{2} Lastly, while the predicted canted-AFM structure is qualitatively consistent with the weak FM observed in the magnetization measurement, we note that future measurement using direct probe techniques such as neutron scattering or muon spin relaxation are needed to confirm the magnetic ground state. 

\begin{table}[h!]
\centering
\begin{tabular}{||c c c c c c c c c c||} 
 \hline
 Element & $S_x$ & $S_y$ & $S_z$  & $L_x$  & $L_y$  & $L_z$  & $M_x$ & $M_y$ & $M_z$ \\ [0.5ex] 
 \hline\hline
 Cu(1) & 0.056 & 0.096 & 0.636 & 0.033 & -0.002 & 0.059 & 0.089 & 0.094 & 0.695\\ 
 Cu(2) & -0.055 & 0.096 & -0.636 & -0.034 & -0.002 & -0.058 & -0.089 & 0.094 & -0.694\\
 Ir(1) & 0.021 & -0.031 & -0.218 & 0.057 & -0.070 & -0.411 & 0.078 & -0.101 & -0.629\\
 Ir(2) & -0.022 & -0.031 & 0.211 & -0.058 & -0.070 & 0.410 & -0.080 & -0.101 & 0.621\\
 \hline
\end{tabular}
\caption{Components of spin $S_{x,y,z}$, orbital $L_{x,y,z}$, and total $M_{x,y,z}$ moments (in the unit of $\mu_B$) in the doubled unit cell are obtained from the GGA+SOC+U calculation for the canted AFM ground state.}
\label{table1}
\end{table}

\section{Conclusions}

We carried out a study of the magnetic and electronic properties of double perovskite (La$_{1-x}$Sr$_x$)$_2$CuIrO$_6$ compounds ($x$ = 0.0, 0.1, 0.2, and 0.3). The undoped La$_2$CuIrO$_6$ undergoes a magnetic phase transition from paramagnetism to AFM at $T_N$ $\sim$ 74 K, followed by a weak FM behavior below $T_C$ $\sim$ 52 K. First-principle density-functional theory (DFT) calculations suggested that a canted AFM order is formed in both the Cu$^{2+}$ and the Ir$^{4+}$ sublattice, which results in the observed weak FM. Two-dimensional magnetism that was observed in many other Cu-based double-perovskites is absent in our samples, indicating the existence of moderate Cu-Ir interaction. DFT calculations of electronic properties provided a strong evidence that La$_2$CuIrO$_6$ is an SOC-driven Mott insulator. Sr-doping decreases magnetic ordering temperatures as well as the electrical resistivity. The high temperature resistivity can be fitted using a variable-range-hopping model, consistent with the existence of disorders in the double-pervoskite compounds.

\acknowledgments

S.X.Z. would like to acknowledge Indiana University (IU) College of Arts and Sciences for start-up support. W.T. acknowledges support from the Youth Innovation Promotion Association, Chinese Academy of Sciences. C.K.L. was supported by the Taiwan Ministry of Science and Technology through Grant No. 03-2112-M-003-012-MY3. H.D.Z. Thanks for the support from NSF-DMR-1350002. We acknowledge the use of facilities at the IU Molecular Structure Center (supported by NSF grant \#CHE-1048613) and in the High Magnetic Field Laboratory, Chinese Academy of Sciences at Hefei. XPS instrument at Nanoscale Characterization Facility of IU Nanoscience center was founded by NSF Award DMR MRI-1126394. C.K.L., P.H.L., J.C.T., and Y.K.W. would like to acknowledge National Center for High-Performance Computing of Taiwan, National Taiwan Science Education Center, Institute of Atomic and Molecular Sciences of Academia Sinica, and M.-Y. Chou for supporting their computer time and facilities.

\end{document}